# Fast Dual-Radio Cross-Layer Handoffs in Multi-Hop Infrastructure-mode 802.11 Wireless Networks for In-Vehicle Multimedia Infotainment

Jayaraj Poroor, Sriram Karunagaran, Sudharsan Sundararajan, Ranjith Pillai
Amrita Research Labs
Amrita Vishwa Vidyapeetham
Kollam, India
{jayaraj, sriramk, sudharsan, ranjith}@arl.amrita.edu

*Abstract*—Minimizing handoff latency and achieving near-zero packet loss is critical for delivering multimedia infotainment applications to fast-moving vehicles that are likely to encounter frequent handoffs. In this paper, we propose a dual-radio cross-layer handoff scheme for infrastructure-mode 802.11 Wireless Networks that achieve this goal. We present performance results of an implementation of our algorithm in a Linux-based On-Board-Unit prototype.

## I. INTRODUCTION & MOTIVATION

Delivering multimedia infotainment applications such as VoIP, IPTV, or drive-by info-fueling to moving vehicles require seamless high-bandwidth connectivity between On-Board Units (OBUs) and Multimedia Services. 802.11 Infrastructure-mode Wireless LAN (WLAN) is a suitable candidate for delivering such services because of its increasingly ubiquitous deployments, availability of cheap commodity off-the-shelf hardware, high bandwidth potential, and strong security. Multi-hop WLANs are better suited for high-bandwidth applications compared to single-hop WLANs (in the latter, each Access Point – AP – must cover a larger area) since the former has enhanced overall capacity through better spatial frequency reuse and provides higher data-rates to mobile nodes (MNs) due to shorter average-distances to APs. Achieving near-zero packet loss during handoff, when vehicle moves from range of one AP to that of another, is critical for delivering multimedia applications to fast moving vehicles, since they are likely to encounter frequent handoffs. Though a number of solutions have been proposed for fast handoffs in WLANs, most of the existing solutions focus either on Layer-2 [1,2] or Layer-3 [3,4] handoff issues without considering cross-layer issues. The necessity for cross-layer approach is discussed in [5], which describes a scheme using single-radio. In this paper we propose a fast cross-layer handoff scheme using dual-radio. Our scheme achieves near-zero packet-loss necessary for delivering multimedia applications to fast-moving vehicles in a multi-hop BSS WLAN environment, which is difficult to achieve in single-radio architectures.

## II. ASSUMPTIONS & TERMINOLOGY

In order to achieve near-zero packet loss during handoff, we make two assumptions which are valid in our application context: (a) Our scheme requires MN to have two radios and requires it to run custom handoff software agent. For an On-Board-Unit (OBU) this is not difficult to realize since it does not have energy and size constraints like a mobile phone. (b) During handoff we only attempt to establish route between the MN and a central Gateway (G), not between two MNs. Since for in-vehicle infotainment, vehicle-to-vehicle communication is not important, this assumption has no limiting effect. Fig. 1 shows the topology of our outdoor multi-hop WLAN test-bed which we use to validate our handoff mechanism. An MN with its two radios MN.RADIO1 and MN.RADIO2 is shown.

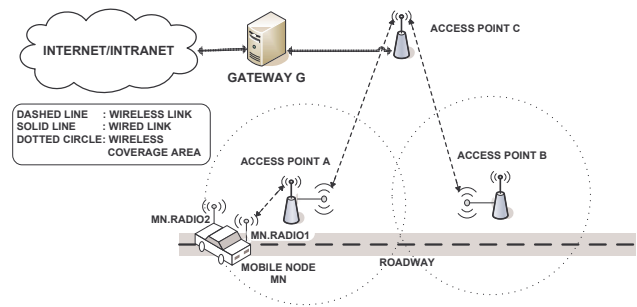

Figure 1. Topology of Outdoor Handoff Testbed

We distinguish between Edge APs to which MNs associate, and Core APs which is only involved in operation of the wireless backhaul. The figure shows two Edge APs - A and B, and one Core AP – C. We assume that MNs communicate with multimedia services via G, which will be typically connected to one or more Core APs via wired links. Therefore, to deliver continuous multimedia services to vehicle, we need to maintain continuous Layer-3 connectivity between G and MNs as MNs roam in the wireless coverage area. Apart from MNs, the G and Edge APs, which are part of the infrastructure, must also run custom handoff agents. We make no assumptions regarding AP-to-AP communication and we employ UDP-based messaging for handoff – thus allowing our system to work well in all 802.11 environments, including the recently standardized 802.11s.

This work has been supported by a grant titled "Intelligent & Interactive Telematics using Emerging Wireless Technologies for Transport Systems", from Technology Information Forecasting and Assessment Council (TIFAC), India

## III. FAST DUAL-RADIO CROSS-LAYER HANDOFF

In a multi-hop network, handoff consists of two components: (a) Layer-2 Handoff – wherein the MN establishes wireless link to a new AP and drops the current link, (b) Layer-3 Handoff – wherein a new route is established to and from the MN via the new AP, while the current route is dropped. The key performance parameters are: (1) Handoff latency and (2) incurred packet-loss. We attempt to minimize (1) while achieving a near-zero (2). Layer-2 Handoff latency is dominated by the delay involved in AP scanning [6]. To eliminate this, we employ a dual-radio mechanism similar to [7,8], wherein one radio (called secondary radio) is always actively scanning for nearby APs while the other one (called primary radio) is associated with an AP and engaged in data transfer. When MN finds that the link quality (LQ) (based on Signal-to-Noise Ratio and Received Signal Strength) of the AP associated to primary radio is below a set threshold and secondary radio has found an AP with better link quality, then secondary radio associates with this new AP, while primary radio maintains its current association. Layer-3 handoff latency is primarily composed of the delays involved in route-discovery, which we minimize by employing GRE-tunnels between G and Edge APs through which G tunnels packets meant for an MN to its associated Edge AP. Therefore, when a handoff occurs, an Edge AP need to communicate this information to G alone and need not propagate it to other nodes, minimizing the route discovery delay. Unlike [3,4] our tunnels are pre-configured, eliminating setup delays during handoffs. In Mobile IP [9] terminology, G acts as Home Agent and the Edge AP acts as Foreign Agent, however to accommodate our cross-layer approach we have not used standard Mobile IP.

Before Layer-2 handoff to a new AP is committed, MN must look at the path-quality from the AP to G in terms of available bandwidth [5]. To give a boundary example, the new AP may provide very good link quality, however its backhaul link may be broken. In this case if Layer-2 handoff is completed without taking Layer-3 path-quality into consideration, packet loss will result. This necessitates a cross-layer approach since the MN can get a commitment for the required minimum path-quality from a new AP only after it has associated with the AP. Moreover, handoff to a new AP must be completed only if its layer-3 path-quality is acceptable to MN. In order to maintain continuous layer-3 connectivity with external nodes, the IP address of MN seen by the outside world must remain same when primary and secondary radios switch. Therefore, we assign a radio-independent Virtual IP address (VIP) to the MN. To outside nodes, both radios of the MN appear to be bound to VIP, though they have auto-configured static private IP addresses (which are never seen by outside nodes). We call this IP Address Mirroring, which is achieved on outbound packets by using Source Network-Address-Translation (SNAT) to rewrite the IP address on outbound packets on both radios to the VIP. IP Address Mirroring is achieved on inbound packets as follows: During handoff process, when MN's secondary radio R associates with AP X, an entry is added to ARP cache of X that map MN's VIP to radio R's MAC Address. This ensures that an associated AP can send packets destined for VIP directly to either of the radio interfaces of MN. Only at the end of the handoff cycle does the primary radio dissociate from its AP, after which it becomes the secondary radio and starts scanning. By employing dual-radios that mirror the same VIP and making sure that the existing primary radio dissociates from its AP only at the end of a handoff cycle (by which time routes are already set up for the other radio), we are able to achieve near-zero packet loss. Fig. 2 shows our handoff-algorithm in operation for the topology depicted in Fig. 1. The figure shows handoff agents - MN.HA, B.HA, and G.HA - coordinating together to execute message-exchanges and primitives of the handoff-algorithm. In Fig. 2, names followed by a set of braces represent primitives e.g. CREATE-ROUTE(..), otherwise names represent messages exchanged. Initially MN.RADIO1 is the primary radio, associated with Access Point A (not shown in Figure 2). As MN moves towards Access Point B, it finds that link quality of RADIO1 (associated with A) is degrading and discovers B to be the better AP in vicinity based on LQ. MN then associates RADIO2 with B and begins the handoff process. The handoff gets completed when MN creates the default outbound route via RADIO2, making it the primary radio and dissociates RADIO1 from A, making it the secondary radio. Messages and primitives for Figure 2 are described in Table 1 and Table 2. Our primitives may be easily realized using standard system calls.

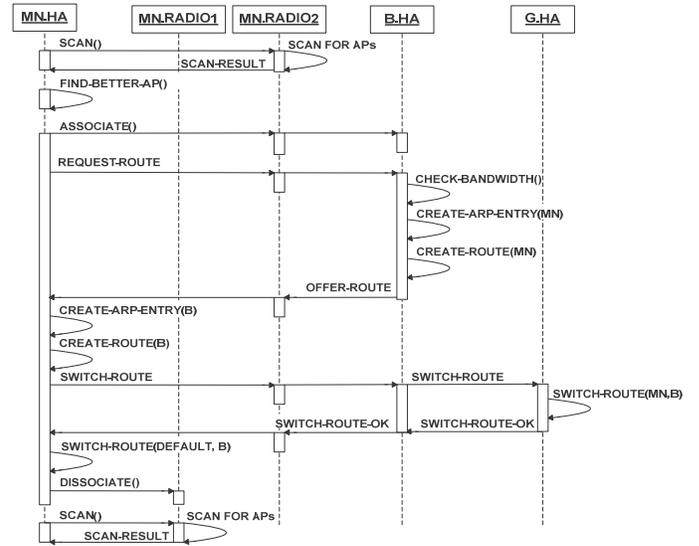

Figure 2. Fast Dual-Radio Cross-Layer Handoff Algorithm

TABLE I. MESSAGES

| Message | Fields (shown capitalized)/Description |
|---|---|
| REQUEST-ROUTE | REQUESTED-BANDWIDTH, MN.RADIO2.MAC-ADDR, MN.FLOATING-IP-ADDR. This message is broadcasted via the secondary radio immediately after it associates with B (new AP). |
| OFFER-ROUTE | AVAILABLE-BANDWIDTH, B.IP-ADDR, B.MAC-ADDR. MN will continue handoff only if B has enough AVABILABLE-BANDWIDTH. |
| SWITCH-ROUTE (MN TO B) | MN.FLOATING-IP |
| SWITCH-ROUTE (B TO G) | MN.FLOATING-IP, B.HOSTNAME |
| SWITCH-ROUTE-OK (G TO B AND B TO MN) | MN.FLOATING-IP, B.HOSTNAME |

TABLE II.     PRIMITIVES

| Primitive | Description |
|---|---|
| SCAN() | Performs scan on the secondary radio by actively sending probe requests on different channels. |
| FIND-BETTER-AP() | This primitive maintains history of recent scans and computes LQ for APs in vicinity, as well as LQ for the AP A to which RADIO1 (primary radio) is associated. When it finds that the LQ of A is degrading below a threshold and that there is an AP (B) that has better LQ, it returns B.BSSID. |
| ASSOCIATE() | Associates RADIO2 (secondary radio) to B |
| CHECK-BANDWIDTH() | B monitors the available bandwidth of the path from B to G. When B gets REQUEST-ROUTE message from MN, it checks if its available bandwidth > REQUESTED-BANDWIDTH. If so, it commits the REQUESTED-BANDWIDTH to MN. The effective available bandwidth for B = Monitored Available Bandwidth – Sum of all committed bandwidths. Within a timeout period, if MN does not complete handoff with B, it will free the committed bandwidth for MN, to prevent bandwidth-leakage. |
| CREATE-ARP-ENTRY(MN) | B creates an ARP entry in its local ARP cache mapping MN.FLOATING-IP-ADDR to MN.RADIO2-MAC-ADDR (provided by MN in REQUEST-ROUTE message). This avoids an extra step required to perform ARP query and will be removed after MN dissociates from B. |
| CREATE-ROUTE(MN) | B creates a route for MN.FLOATING-IP-ADDR via its radio-interface to which MN is associated (B.AP-RADIO). After MN dissociates from B, this route will be removed. |
| CREATE-ARP-ENTRY(B) | MN creates an ARP entry in its local ARP cache mapping B.IP-ADDR to B.MAC-ADDR (provided by B in OFFER-ROUTE message). This avoids an extra step required to perform ARP query and will be removed after MN dissociates from B. |
| CREATE-ROUTE(B) | MN creates a route for B.IP-ADDR via MN.RADIO2 After MN dissociates from B, this route will be removed. |
| SWITCH-ROUTE(MN,B) | G switches the route for MN.FLOATING-IP-ADDR to the virtual interface corresponding to GRE tunnel for B. This switches the inbound route for MN from A to B. |
| SWITCH-ROUTE(DEFAULT, B) | MN switches the DEFAULT route via its RADIO2 interface with B as the router. This switches the outbound route for MN from A to B. |
| DISSOCIATE() | MN dissociates MN.RADIO1 from A, after which MN.RADIO1 becomes the secondary radio used for scanning. |

## IV. IMPLEMENTATION & DISCUSSION

Similar to the indoor experimental results of make-before-break algorithm presented in [7], we observed zero packet loss during handoffs in laboratory setup. Our indoor handoff latencies were around 50 ms compared to less than 10 ms latency of [7]. This is primarily owing to the additional messages necessary for ensuring the multi-hop path-quality. For our outdoor test-bed, we implemented our algorithm in an OBU prototype built using Linux PC having two Atheros mini-PCI cards. We installed OBU inside a vehicle with two roof-mounted 5 dBi omni-antennas. We built APs running our HAs using Soekris net4826 Linux Single-Board-Computers. We ran the vehicle through the test-bed given in Fig. 1 at 40 kmph. During each run, we generated 10,000 ICMP echo requests from OBU with 10ms interval. The packet drops and handoff latencies from 10 outdoor experiments are presented in Fig. 3 and Fig. 4, respectively. The average packet-loss in outdoor environment is ~2 packets per 10,000 packets, which is near-zero percent. The outdoor latencies have increased due to the retries owing to the loss of broadcasted REQUEST-ROUTE messages in outdoor environment. The average latency observed is ~80ms, which will require an overlap between APs of only 2.21m at typical highway speed of 100kmph.

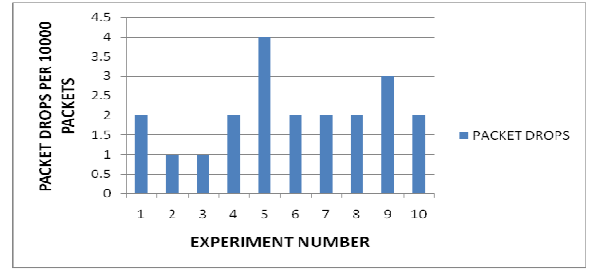

Figure 3.  Fast Dual-Radio Cross-Layer Handoff Algorithm

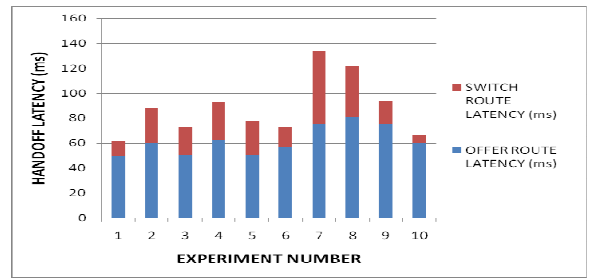

Figure 4.  Handoff latencies during a run of 10 handoffs at 40 kmph

To accommodate the presence of other wireless networks such as WiMax/UMTS, our mechanism may be potentially extended to perform vertical handoffs. We will be investigating this in future work.